\documentclass[12pt,twoside]{article}   %For printing on two sides of page
\usepackage[super,sort,comma]{natbib}
\usepackage{fancyhdr}		%Gives headers and footers defined below
\usepackage[section]{placeins}   %
\usepackage{graphicx}
\usepackage{amsmath,amssymb,amsfonts}
\usepackage{algorithmicx}
\usepackage{textcomp}
\usepackage{algorithm}
\usepackage{algpseudocode}

\makeatletter \renewcommand\@biblabel[1]{$^{#1}$} \makeatother
\newcommand{\note}[1]{\mbox{}\\ \noindent \rule{16cm}{0.5mm} \\
{\em #1} \\ \noindent \rule{16cm}{0.5mm}
\typeout{    }
\typeout{***********note active on this page *************************}
\typeout{Note: #1  }
\typeout{****************************************end Note}
}

\renewcommand{\note}[1]{}

\newcommand{\cen}[1]{\begin{center} #1 \end{center}}

\usepackage{hyperref}
\hypersetup{ colorlinks,
    citecolor=blue,
    filecolor=blue,
    linkcolor=blue,
    urlcolor=blue
}

\usepackage{xcolor}
\definecolor{gray}{rgb}{0.6,0.6,0.6}
\definecolor{red}{rgb}{0.85,0,0}
\definecolor{green}{rgb}{0,0.85,0}
\definecolor{blue}{rgb}{0,0,0.85}
\definecolor{beige}{rgb}{0.92,0.87,0.78}
\usepackage[all]{hypcap}    %causes link to figures to go to figure, not caption
\begin{document}

\cen{\sf {\Large {\bfseries Report on the AAPM deep-learning sparse-view CT (DL-sparse-view CT)  Grand Challenge} \\  
%\vspace*{10mm}
\vspace*{5mm}
Emil Y. Sidky and Xiaochuan Pan
 } \\
Department of Radiology, The University of Chicago, 5841 S. Maryland Ave., Chicago, IL 60637, USA
%\vspace{5mm}\\
\vspace{3mm}\\
Version typeset \today\\
}

\pagenumbering{roman}
\setcounter{page}{1}
\pagestyle{plain}
Author to whom correspondence should be addressed. email: sidky@uchicago.edu\\
% note, probably best not to use a student's e-mail as it won't be valid for
% very long.

\begin{abstract}
\noindent
{\bf Purpose:}
The purpose of the challenge is to find
the deep-learning technique for sparse-view CT image reconstruction that can yield the minimum RMSE under ideal conditions,
thereby addressing the question of whether or not deep learning can solve inverse problems in imaging.\\
{\bf Methods:}
The challenge set-up involves a 2D breast CT simulation, where the simulated breast phantom has random
fibro-glandular structure and high-contrast specks. The phantom allows for arbitrarily large training
sets to be generated with perfectly known truth. The training set consists of 4000 cases where each
case consists of the truth image, 128-view sinogram data, and the corresponding 128-view 
filtered back-projection (FBP) image.
The networks are trained to predict the truth image from either the sinogram or FBP data. Geometry
information is not provided. The participating algorithms are tested on a data set consisting of 100
new cases.\\
{\bf Results:} 
About 60 groups participated in the challenge at the validation phase, and 25 groups submitted test-phase
results along with reports on their deep-learning methodology. The winning team improved reconstruction
accuracy by two orders of magnitude over our previous CNN-based study on a similar test problem.\\
{\bf Conclusions:} 
The DL-sparse-view challenge provides a unique opportunity to examine the state-of-the-art in deep-learning
techniques for solving the sparse-view CT inverse problem.
\end{abstract}
%\note{This is a sample note.}

\newpage     %may or may not be needed

%The table of contents is for drafting and refereeing purposes only. Note
%that all links to references, tables and figures can be clicked on and
%returned to calling point using cmd[ on a Mac using Preview or some
%equivalent on PCs (see View - go to on whatever reader).
\tableofcontents

\newpage

\setlength{\baselineskip}{0.7cm}      %double spacing		

\pagenumbering{arabic}
\setcounter{page}{1}
\pagestyle{fancy}

\section{Introduction}
\label{sec:introduction}

The American Association of Physics in Medicine (AAPM) sponsored a recent Grand Challenge that
addresses the use of deep-learning (DL) and convolutional neural networks (CNNs) for solving
the inverse problem associated with sparse-view computed tomography (CT) image reconstruction \cite{aapm1,aapm2,dlsparse}.
The challenge began on March 17th 2021, when the training data became available, and it concluded on June 1st 2021,
the last day for submission of test results. During the challenge there were approximately 60 teams that were
actively participating and around 500 trial submissions were made. In the final test phase, 25 team submitted their
image predictions along with a report on their DL methodology. This article presents the background and
results of the challenge along with an analysis on what we have learned about the use of DL for solving inverse
problems in tomographic imaging.

The idea for the challenge started from claims made in the literature that
CNNs can solve inverse problems that arise in CT image reconstruction \cite{jin2017deep,han2018framing,zhu2018image}.
In response to those claims, we performed a study along with Lorente and Brankov \cite{sidky2020cnns},
where we investigated the question of whether or not CNNs can solve the inverse problem associated with sparse-view 
CT image reconstruction.
In our article \cite{sidky2020cnns}, we attempted to implement CNN-based image reconstruction based
on \cite{jin2017deep,han2018framing} and we were not able to demonstrate that CNNs can solve the sparse-view CT problem.
The conclusion of our work was that there is no evidence that CNNs can solve such problems despite the
claims made in the literature. We did, however, acknowledge that there may be other DL-based image reconstruction
algorithms that are capable of inverting the sparse-view CT model.

Carrying out a broad survey of DL-based methods for CT image reconstruction, is the motivation for the DL-sparse-view
CT challenge. In effect the challenge is a ``crowd-source'' effort where researchers could apply their own
DL technology on a well-defined sparse-view CT problem. The goal of the challenge is accurate recovery
of the test images from ideal noiseless projection data; accordingly algorithms can be ranked unambiguously with
root-mean-square-error (RMSE) and the floor of this metric is zero at which point one can say that the
CT inverse problem is solved.

We are happy to report that the participation and the ingenuity of the developed DL-based algorithms greatly
exceeded our expectations.
The winning team of DL-sparse-view CT challenge report a RMSE score -- two orders of magnitude smaller than what was reported
for the DL-based method in Sidky {\it et al.}\cite{sidky2020cnns}
for a similar image reconstruction set-up. Furthermore, there is a healthy
spread of scores resulting from the numerous DL-based implementations employed. As such, it is possible to start to gauge
the success of various DL designs for CT image reconstruction.

In Sec. \ref{sec:methods} the details of the DL-sparse-view CT challenge are outlined, discussing the problem
set-up, evaluation metrics, and challenge logistics. In Sec \ref{sec:results}, the results of the challenge are
presented, including statistics of all scores along with groupings of scores according to the general class
of DL methodology and a focus on the results of the top five participating teams. Section \ref{sec:discussion}
discusses how the results relate to the question of solving the CT inverse problem and concludes the paper.

\section{Methods}
\label{sec:methods}
\subsection{Inverse problem theory}
\label{sec:ip}

%The recent focus on data driven methods has side-lined inverse problem theory as being esoteric and irrelevant,
%and one of the motivations of the challenge is to renew interest in performing such studies as they have
%tangible practical benefit.
%Hundreds of DL-based image reconstruction algorithms have
%appeared in the literature over the last five years, yet there is little organization to the generated material.
%Which algorithm should be applied in which situation? How do we design and specify the network parameters?
%How good are these algorithms at accomplishing a desired imaging task? 
%One avenue of beginning to answer
%these questions, is to study DL image reconstruction in controlled simulations that address
The motivation for the DL sparse-view CT challange is the investigation of the inverse problem associated
with sparse-view CT image reconstruction, and accordingly we provide a brief review of inverse problem theory.
At a high level, inverse problems require the specification of a measurement model \cite{bal2012introduction}
%\begin{linenomath}
\begin{equation*}
y = \mathcal{M}(x),
\end{equation*}
%\end{linenomath}
where $x$ represents unknown model parameters; and $\mathcal{M}$ is an operator that yields the data $y$ from $x$.
For deriving inverses, this measurement model usually represents a simplified physical model, which is on the one
hand complex enough to capture the dominant physics of the imaging problem but simple enough that deriving an inverse
is a tractable mathematical problem.
For the sparse-view CT problem of interest, $x$ is a discrete CT image; $\mathcal{M}$ represents discrete-to-discrete
linear forward projection; and $y$ is the sparsely sampled sinogram.
Solving the inverse problem entails deriving $\mathcal{M}^{-1}$ such that 
%\begin{linenomath}
\begin{equation*}
x = \mathcal{M}^{-1}(y).
\end{equation*}
%\end{linenomath}
Solving an inverse problem is a binary issue; either test images are recovered exactly for this model
or they are not.

Inverse problem studies confer generalizability to a proposed model inversion technique.
Once it is established that $\mathcal{M}^{-1}$ inverts the measurement
model, we know that any $x$ can be recovered from data $y$ in the range of $\mathcal{M}$.
Also, generalizable stability results rely on demonstration of inverting the idealized model.
A realistic, perhaps intractable model $\mathcal{M}_\text{phys}$,
can be written in terms of an idealized model as
%\begin{linenomath}
\begin{equation*}
\mathcal{M}_\text{phys}(x) = \mathcal{M}(x) + \epsilon(x),
\end{equation*}
%\end{linenomath}
where $\epsilon(x)$ represents both deterministic and statistical model error.
A stability result puts a bound on the reconstruction error $\mathcal{M}^{-1}(\epsilon(x))$
only if $\mathcal{M}^{-1}$ is the true inverse of the idealized model $\mathcal{M}$.

Inverse problem studies also play a central role in the reproducability of image reconstruction algorithms.
From an engineering perspective, all image reconstruction algorithms are implemented on a computer,
and the more complex an algorithm is, the more difficult it is to debug.
Checking that $x$ can be recovered accurately from ideal data $y$
in the range of $\mathcal{M}$ provides a stringent test on the computer code.
If the model inverse $\mathcal{M}^{-1}$ is known, any error in the recovery of
$x$ must be due to implementation or numerical error.
Note that applying a model inverse to noisy corrupted data will yield
ambiguous results with respect to algorithm validation. There will be error in $x$, but it is difficult
to say whether this error is due to implementation error, data error, or a combination of both.

The present DL-sparse-view CT challenge is simulation-based and it addresses the concrete issue as to
whether or not a deep-learning network can solve the inverse problem of sparse-view CT image reconstruction.

\subsection{Sparse-view CT simulation and object modeling}

The simulation is based on breast CT and it is similar to the one used in
Sidky {\it et al.}.\cite{sidky2020cnns}
The CT data model is the standard discrete-to-discrete linear model that is commonly used in CT iterative
image reconstruction
%\begin{linenomath}
\begin{equation*}
g=Xf,
\end{equation*}
%\end{linenomath}
where $g$ represents the discrete sinogram data; $X$ is a linear transform encoding X-ray projection; and $f$
is the pixelized image array representing the scanned subject. Matching up with the inverse problem formalism,
the image pixels $f$ represent the unknown model parameters $x$; the measurement model $\mathcal{M}(\cdot)$ is
matrix multiplication by $X$; and the sinogram data $g$ are associated with the the model measurements $y$.
The image array is 512x512 pixels covering an area (18cm)$^2$. The line-intersection method is used for generating
the matrix elements of $X$.
The scan configuration is 2D sparse-view CT, where 128 evenly space projections are acquired over a 360 degree
scan. A fan-beam configuration is used with a source-to-detector distance of 100 cm and a source-to-center-of-rotation
distance of 50 cm. The linear detector array has 1024 detector elements.
For the challenge image reconstruction problem, the data $g$ are generated by applying $X$ to test images $f$
and the goal is to see if participants can recover $f$ from the sinogram $g$. As the image size is 512$^2$
and the sinogram size is 128$\times$1024, it is clear that the problem is under-sampled because there are more
unknown pixel values than sinogram measurements. 

\begin{figure}[!t]
\centerline{\includegraphics[width=1.0\columnwidth]{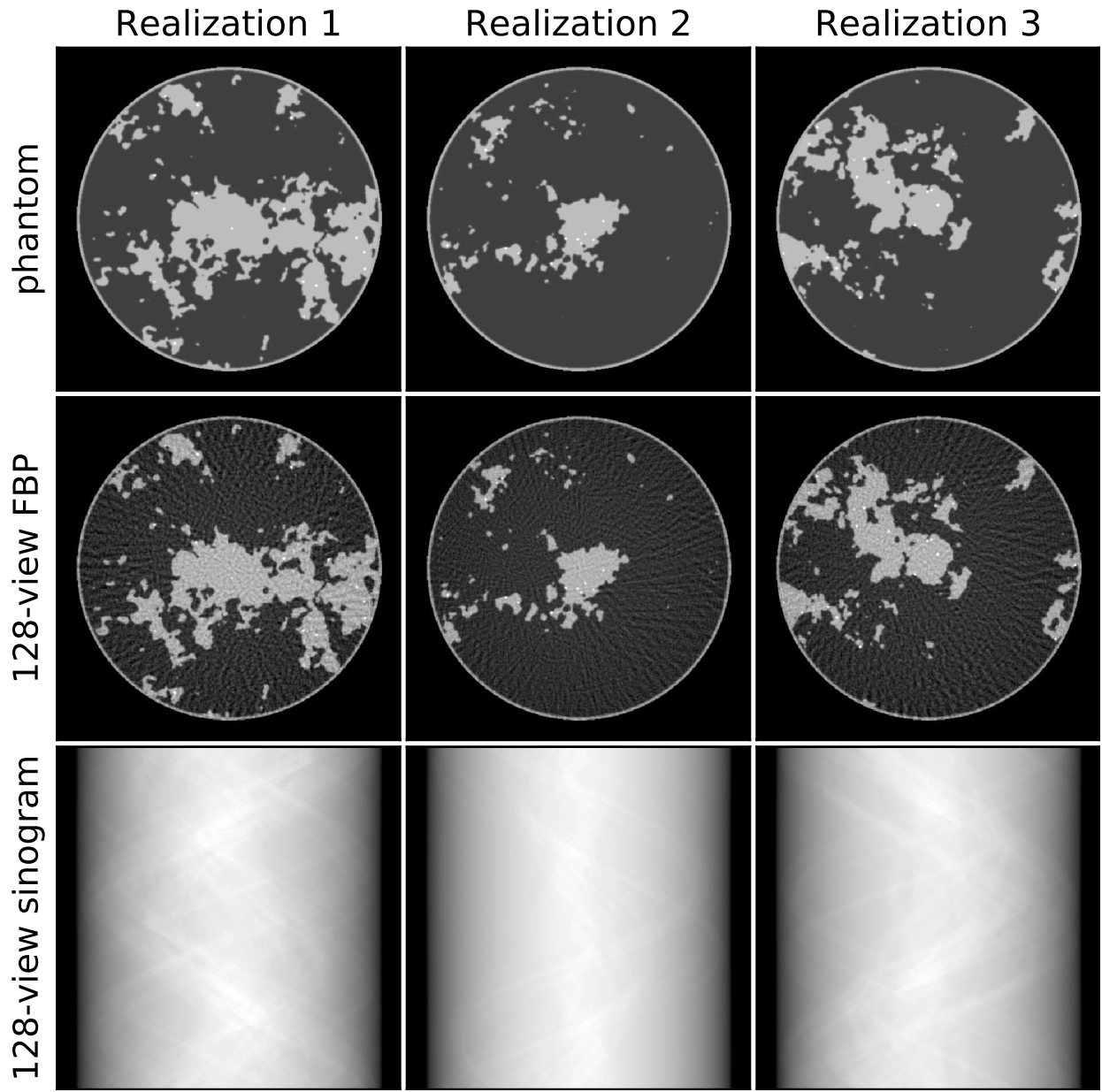}}
\caption{(Top row) Three realizations of the breast phantom model. (Middle row) Corresponding FBP image
obtained from the 128-view sinogram (Bottom row). The images are shown in a gray scale window
of [0.174,0.253] cm$^{-1}$.
}
\label{fig:phantoms}
\end{figure}

Because the image reconstruction algorithms in question are data-driven techniques, the object modeling
is the most crucial component to the simulation. The breast phantom model \cite{Reiser10,sidky2020cnns} is stochastic so that multiple
realizations can be generated for the purpose of training, validation, and testing.  Three realizations
of this object model are shown in Fig. \ref{fig:phantoms} along with the corresponding sinograms
and filtered back-projection (FBP) images. All breast slice realizations are centered
on the middle of the image area, and have a circular profile of radius 8 cm. Four tissues are modeled:
adipose with an attenuation of 0.194 cm$^{-1}$, skin and fibroglandular tissue at 0.233 cm$^{-1}$, and 
single-pixel specks resembling microcalcifications with variable attenuation ranging from 0.333 to 0.733 cm$^{-1}$.
The random components of the phantom are the fibroglandular morphology, the speck number (between 10 and 25 per phantom
realization), speck location, and speck amplitude.  The speck location is restricted to the fibroglandular regions.
After generating the phantom with discrete gray levels, we apply Gaussian smoothing with a full width at half maximum
of one pixel, in this way the gray levels of the phantom varies smoothly between the tissue types.

The breast phantom images are challenging to recover mainly due to the complex fibroglandular morphology.
The image smoothing and random speck amplitudes prevents hard-coding discrete gray levels or a tight upper
bound for the fibroglandular component of the phantom. Despite the object complexity, it is amenable to
sparsity-exploiting image reconstruction techniques because the phantom can be expressed as the gaussian
blurring of an image that has a sparse spatial gradient magnitude image. In Sidky {\it et al.}\cite{sidky2020cnns}, 
accurate image reconstruction for this CT configuration and breast model is demonstrated with a standard
sparsity exploiting image reconstruction technique. The purpose of designing the object so that 
model inversion is possible by exploiting gradient-sparsity is to demonstrate that solution of this
sparse-view CT system is possible. In this way, for whatever image reconstruction technique that is developed,
we know in advance that numerically exact object recovery is possible.

\subsection{Challenge logistics and scoring metrics}

In order to explain the information provided for this challenge, we refer to the taxonomy of
DL techniques for inverse problems in imaging provided in Ongie {\it et al.}.\cite{ongie2020deep}
The DL taxonomy is organized along two axes: (1) knowledge of the system matrix, and (2)
quality of the training data. Because we are interested in image recovery under ideal conditions, we provide
training data of the highest quality, i.e. supervised learning with exact knowledge of the ground truth
images. The DL sparse-view CT challenge focuses on the first axis of the DL taxonomy: ($i$) $X$ is known
exactly, ($ii$) $X$ is known only at test time, ($iii$) $X$ is partially known, and ($iv$) $X$ is unknown.
In the literature for DL-based image reconstruction for CT, by far the most common approach involves
predicting the truth image from an FBP reconstructed image, using a U-net to remove
the streak artifacts inherent in the latter.\cite{jin2017deep,han2018framing}
This strategy falls into the first category of DL techniques because the exact geometry of the
scan configuration, which specifies $X$, is exploited in obtaining the FBP image that serves as the
input to the CNN. On the other extreme, the automated transform by manifold approximation (AUTOMAP)
framework\cite{zhu2018image} is an example of the last category; for this
type of approach the transformation from sinogram to image is learned without knowledge of $X$
or the scan geometry.
There are also numerous techniques for incorporating a CNN into an iterative procedure,
see Ongie {\it et al.}\cite{ongie2020deep} for a survey. The most common form of this approach
is the unrolled iterative technique, which
belongs to the first category where knowledge of $X$ is available.
In the unrolled iterative technique a finite number of iterations are performed, alternating
between a regularization step and a standard data-consistency step. The CNN is used for the
regularization step, and knowledge of $X$ is exploited in the data-consistency step.
Finally, standard iterative image reconstruction can be used with inexact knowledge
of $X$, i.e. category ($iii$). The resulting images will have artifacts resulting from the
use of an inexact system matrix $X$, and these artifacts can be mitigated by a U-net CNN in
a similar way as is done with the FBP-CNN method.

For the DL sparse-view CT challenge, we provided 128-view FBP images, 128-view sinograms, and
the knowledge that the scan geometry is a circular, fan-beam configuration. We did not provide
the geometry parameters or the scheme for generating the weights of the system matrix $X$.
Under these conditions, DL methods from category ($i$) had to involve the use of the
128-view FBP images. The challenge does not test category ($ii$) methods because we used the same
$X$ in generating the training and testing data. Category ($iii$) methods could be developed for the
challenge by exploiting a low-parameter description of $X$ and using the sinograms and truth images
in the training set to arrive at scan parameter values for an estimate of $X$. Finally, the
challenge also admitted category ($iv$)
approaches which do not require $X$, such as AUTOMAP, or that estimate
the scan geometry parameters simultaneously with the image itself. By not providing the
system matrix $X$, the challenge barred hybrid DL/iterative appoaches in category ($i$).
This was necessary, because the design of the image reconstruction problem is such that it can
be solved by constrained, TV-minimization with the smoothed-object model.\cite{sidky2020cnns}
With knowledge of $X$, this TV-minimization algorithm could be used to obtain the truth image from the given
test set sinograms.

To kick off the DL-sparse-view CT challenge 4,000 training cases were released on March 17th 2021 \cite{dlsparse}.
Each training case consisted of the truth image, the 128-view sinogram, and a 128-view
FBP, and example cases are shown in Fig. \ref{fig:phantoms}. The 128-view
FBP images were generated by ramp-filtering the 128-view sinogram and applying back-projection.
%The goal of the network training is to obtain a CNN or CNN-based algorithm can that predict the
%truth image from input data either in the form of
%the 128-view sinogram or the 128-view FBP image.
The back-projection implemented for the FBP image
uses the exact transpose of the projection matrix $X$
so that use of the 128-view FBP image in the DL methodology would constitute
a category ($i$) technique. Using the matrix transpose of the line-intersection
method, also known as ray-driven projection, is non-standard for FBP implementation
in practice. For producing
high quality FBP images, with fine angular sampling, the standard back-projection implementation
is pixel-driven which is essentially a trapezoid rule discretization of the continuous back-projection
integral. This issue of mis-matched projector/backprojector pairs has been studied in the context
of iterative image reconstruction \cite{zeng2000unmatched,lorenz2018randomized}.
In developing our CNN implementation \cite{sidky2020cnns}, we found that use of $X^\top$ as the back-projector
yielded more accurate CNN training than use of pixel-drivel back-projection. Pixel-driven back-projection
does involve slight smoothing of the sinogram data as the values that are back-projected need to be interpolated
from the available samples. We hypothesize that this slight smoothing does make the image recovery problem
a little more difficult because it would change the recovery problem from category ($i$) (exact knowledge of $X$)
to category ($iii$) (approximate knowledge of $X$).
% In any case, the precise back-projection algorithm is an important implementation
%issue that needs to be further explored, but for the purpose of simplicity we only provided data for
%the stated 128-view FBP implementation.

%The two types of input data are provided to accommodate different CNN-based image reconstruction strategies.
%The artifact removal strategy involves some form of
%U-net that takes the 128-view FBP image and attempts to predict the truth image \cite{jin2017deep,han2018framing}. End-to-end
%networks take the sinogram data as input. There are also unrolled iterative techniques,
%which use a trained network as a regularizer, but we did not provide the precise geometry of the scan
%so such an approach would need to discover the scan geometry from the available image and sinogram data.

Participants could begin exploiting the 4000-case training data set for developing the various CNN-based approaches
on March 17th. On March 31st, a validation set consisting of 10 cases without the truth was released. Participants
could submit their predictions for the 10 cases to the challenge website and have their results scored an unlimited
number of times. They
could decide whether or not to submit their scores to the validation-phase leaderboard. In this way, participants
could compete openly or decide to keep their results secret. The final test phase began on May 17th with the
release of 100 new cases without truth, and participants were only allowed to submit three times. The competition
ended May 31st, and final
scores for the challenge were taken as the best of the three submissions.

For scoring the submissions, two RMSE-based metrics were used.
The primary metric that determined the participant ranking was the RMSE averaged over all 100 test case predictions
%\begin{linenomath}
\begin{equation*}
s_1 = \frac{1}{100} \sum_{i=1}^{100} \sqrt{ \frac{ \|t_i-r_i\|^2_2 } {n} },
\end{equation*}
%\end{linenomath}
where $r$ and $t$ are the reconstructed and truth images, respectively; $i$ is the test image index; $n=512^2$ is the number
of pixels in the truth image.
In case of a tie-breaking situation, we also computed a second metric, which is a worst-case ROI RMSE
%\begin{linenomath}
\begin{equation*}
s_2 = \max_{i,c} \sqrt{ \frac{b_c^\top \|(t_i-r_i)\|^2_2 } {m} },
\end{equation*}
%\end{linenomath}
where $b_c$ is the image of an indicator function for a 25x25 pixel ROI centered on coordinates $c$, and $m=625$
is the number of pixels in the test ROI.
As stated earlier, the image reconstruction problem is designed so that it is known that exact recovery of
the truth images is possible. As a result it is possible to drive both $s_1$ and $s_2$ to zero.

The DL sparse-view CT challenge\cite{dlsparse} is hosted by the MedICI Platform
(see \url{https://www.medici-challenges.org/}). The MedICI team hosts the challenge
data, which is accessible by sftp protocol, and they also provided the initial
implementation of the challenge on CodaLab
(see \url{https://github.com/codalab/codalab-competitions/wiki/Project_About_CodaLab}).
Using CodaLab, the MedICI team created the challenge and leaderboard format, set up the submission checking, and
created the python plugins that perform the computation of the challenge metrics. The authors of this
challenge report
were given access to the CodaLab source page in order to edit the text of the DL sparse-view
CT Challenge website. Upon release of the training data, the challenge website ran in a completely
automated fashion allowing for participant submission of training, validation, and test results
in the announced time windows of each phase of the challenge.

\section{Results}
\label{sec:results}

\begin{figure}[!t]
\centerline{\includegraphics[width=0.5\columnwidth]{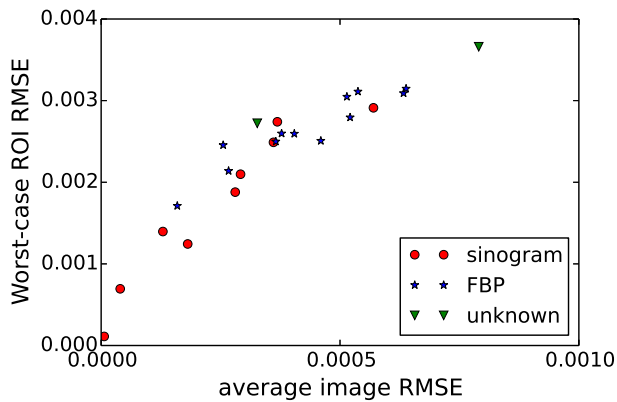}}
\caption{Scatter plot of test phase results with average RMSE $s_1$ on the $x$-axis
and worst-case ROI RMSE $s_2$ on the $y$-axis. The red dots show results for teams
that used the sinogram (and possibly the 128-view FBP) data for training their CNN-based
algorithms. The blue stars indicate results for teams that exploited only the 128-view FBP image
data. The two green triangles are for results that did not specify their methodology.
}
\label{fig:testphaseSC}
\end{figure}

A scatter plot of the final test phase results appears in Fig. \ref{fig:testphaseSC}.
Of the shown results the, the average RMSE $s_1$ ranges from $6.37 \times 10^{-6}$ to
$7.90\times 10^{-4}$ and the worst-case ROI RMSE $s_2$ ranges from $1.11 \times 10^{-4}$
to $3.66 \times 10^{-3}$.
Interestingly, the rank order of $s_1$ is not the same as
that of $s_2$. There is, however,  a healthy spread of values in both $s_1$ and $s_2$, and
it is not necessary to go to tie-breaking rules to determine the challenge leaders and winners,
which is based solely on the ranking of $s_1$. The top five teams are listed in Table \ref{tab:winners}.

\begin{table}
\begin{center}
\begin{tabular}{c| c c} 
 \hline
 Username/team &  $s_1$ & $s_2$   \\ [0.5ex] 
 \hline\hline
 Max/Robust-and-stable  & 6.37$\times 10^{-6}$  & 1.11$\times 10^{-4}$   \\ 
 \hline
 TUM/YM\&RH             & 3.99$\times 10^{-5}$  & 6.95$\times 10^{-4}$   \\
 \hline
 cebel67/DEEP\_UL       & 1.29$\times 10^{-4}$  & 1.39$\times 10^{-3}$   \\
 \hline
 deepx/--               & 1.59$\times 10^{-4}$  & 1.71$\times 10^{-3}$   \\
 \hline
  Haimiao/HBB           & 1.81$\times 10^{-4}$  & 1.24$\times 10^{-3}$   \\ [1ex] 
 \hline
\end{tabular}
\end{center}
\caption{The team memberships in rank order are: (1) Robust-and-stable: Martin Genzel$^a$ (m.genzel@uu.nl),
Jan Macdonald$^b$ (macdonald@math.tu-berlin.de), and Maximillian M\"arz$^b$ (maerz@math.tu-berlin.de);
(2) YM\&RH: Youssef Mansour$^c$ (y.mansour@tum.de), and Reinhard Heckel$^{c,d}$;
(3) DEEP\_UL: C\'edric B\'elanger$^e$ (cedric.belanger.2@ulaval.ca), Maxence Larose$^e$, Leonardo Di Schiavi Trotta$^e$,
R\'emy B\'edard$^e$, and Daniel Gourdeau$^e$; (4) deepx: Yading Yuan$^f$ (yading.yuan@mountsinai.org);
(5) HBB: Haimiao Zhang$^g$ (hmzhang@bistu.edu.cn), Bin Dong$^h$, and Baodong Liu$^i$.
The participating institutions are: $^a$Utrecht University, $^b$Technical University of Berlin, $^c$Technical University of Munich,
$^d$Rice University, $^e$Universit\'e Laval, $^f$Icahn School of Medicine at Mount Sinai, $^g$Beijing Information Science and Technology University,
$^h$Peking University, $^i$Chinese Academy of Sciences.
\label{tab:winners}}
\end{table}

The plot in Fig. \ref{fig:testphaseSC} also shows the breakdown of scores for participants
that exploited only the 128-view FBP images (blue stars) and those that also made use of the 
sinogram data (red circles). It is clear that use of the sinogram data helps to improve the
image reconstruction accuracy as the red dots cluster at lower values for both $s_1$ and $s_2$ metrics;
moreover the advantage of exploiting the sinogram persists even though these methods
only had approximate knowledge of the system matrix $X$ while the FBP image are created with
the exact system matrix transpose $X^\top$.
Nevertheless the spread in both the red and blue clusters is large suggesting that implementation
details for both approaches matter a great deal.
For participants who exploited the sinogram data, the majority used this data to find the unknown
scan parameters of the fan-beam acquisition geometry thus estimating the physical forward model of
the imaging system. Even with such knowledge, the under-sampling of the scan is still a significant
barrier to obtaining exact image recovery. The participants using the physical forward model
all developed hybrid schemes that combined physical data consistency with CNN-based image regularization.

The top five performing algorithms listed in Table \ref{tab:winners} emphasize the success of the DL-sparse-view
challenge in catalyzing CNN-based algorithm development for CT image reconstruction. For reference, our
implementation of the U-net methods of Jin {\it et al.}\cite{jin2017deep} and Han and Ye\cite{han2018framing},
which exploit only the 128-view FBP image data, yielded an average RMSE of 6.76$^{-4}$ cm$^{-1}$
with the caveat that there are differences between the two studies; for the challenge the scan configuration
is fan-beam instead of parallel-beam, high-contrast specks have been added to the phantom, and the test set
consists of 100 images instead of 10. Nevertheless, the amount of training data is 4,000 cases in both our
study and the DL-sparse-view CT challenge, The winning team, Robust-and-stable, improved on our mark by two orders
of magnitude! The fourth place finisher, deepx, achieved the best results among all teams using only the 128-view
FBP data, and the deepx result improves over our average RMSE by nearly a factor of 4.

\begin{figure}[!t]
\centerline{Truth ROI~~~~~~Predicted ROI~~~~~~~~Difference}
\centerline{\includegraphics[width=0.55\columnwidth]{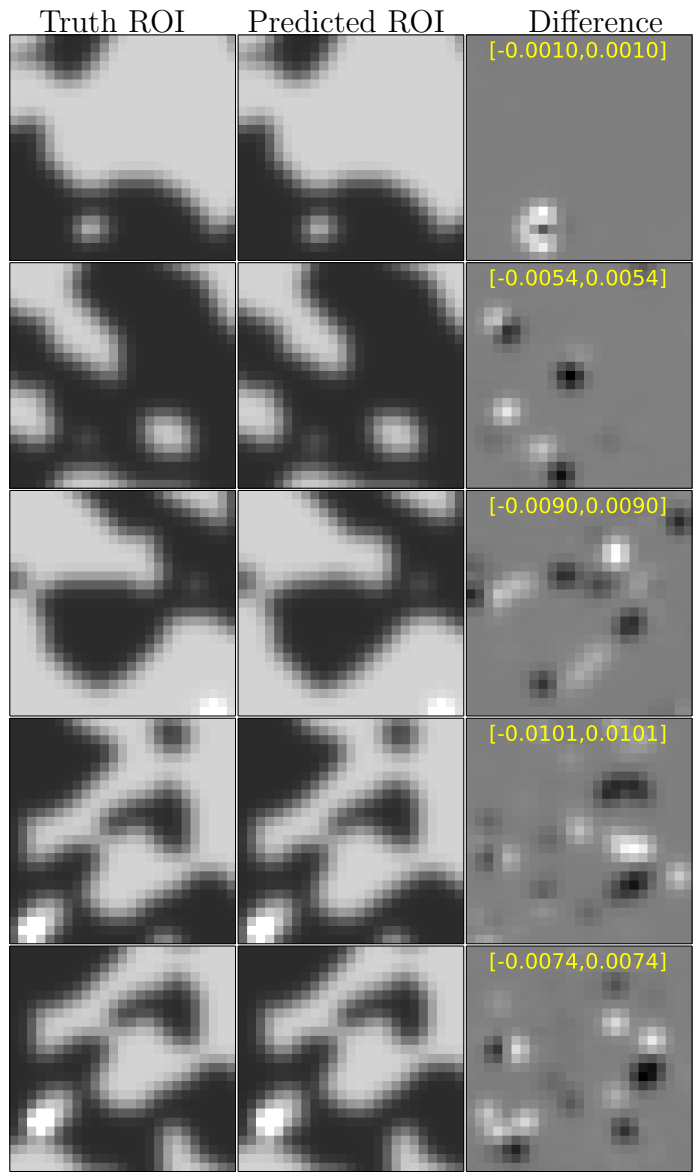}}
\caption{Worst case ROIs for the top five performing teams where the order
from top to bottom corresponds to the rank order of average RMSE $s_1$ from
lowest to highest. The shown ROI is the one that yielded the worst-case ROI RMSE $s_2$
for the respective algorithms. The truth and predicted ROIs are shown in a gray
scale [0.184,0.243] cm$^{-1}$. The ROI difference gray scale is set to $\pm$ the maximum
pixel error in the ROI, and the numerical values are shown in the difference images
in units of cm$^{-1}$. For reference, the contrast between adipose and fibro-glandular
tissue in the phantom is 0.039 cm$^{-1}$. The locations of the worst-case ROIs
from top to bottom are: (262,135) of image realization 39, (275,318) of image realization 26,
(284,306) of image realization 26, (242,340) of image realization 26, and (246,339)
of image realization 26; (0,0) and (512,512) are the upper-left and lower-right corners of the image,
respectively.
}
\label{fig:ROIs}
\end{figure}

The excellent performance of the top five algorithms is demonstrated in the worst-case ROI images shown in Fig. \ref{fig:ROIs}.
In each case, the shown 25x25 ROI is the one with the largest RMSE discrepancy between truth and prediction, and in each
case it is difficult to detect visual differences between true and predicted ROIs. The discrepancy is only noticeable in 
the corresponding ROI difference image, where the gray scale is centered on zero and ranges to the maximum pixel difference
in the ROI. The fifth place team, HBB, achieved a maximum pixel error of 0.0074 cm$^{-1}$, which is a factor of five smaller
than the tissue contrast between the adipose and fibro-glandular tissue. 

There are a few other points of interest to mention about the worst-case ROI results. The fourth place result has a higher $s_2$
than the fifth place result, and more generally, we note that it is particularly difficult for approaches, which only exploit
the 128-view FBP data, to drive the $s_2$ metric to zero. Four of the five top place finishers had their worst-case ROI appear
in the 26th image; only the first place finisher had their worst-case ROI occur in the 39th image of the 100-case test set.
The fourth and fifth place finisher even had their worst-case ROI appear nearly at the same location as their corresponding
ROIs are only shifted from each other by a few pixels. This is particularly surprising because deepx uses only the 128-view FBP
data, while HBB exploits the sinogram data.

We briefly summarize the methodologies of the top five performing teams, leaving the participants the option to
publish their own detailed manuscripts on their methodology:\\
The winning team, Robust-and-stable, estimated the system geometry from the 128-view training sinograms
and truth images. Then, they made use of only the 128-view sinogram data along with the estimated system geometry in testing.
%The winning team, Robust-and-stable, developed an approach that made use of only the 128-view sinogram data.
%They estimated the system geometry from the training sinograms and truth images.
The system geometry was used for a
data-consistency step in an iterative scheme and for generating their own 128-view FBP images.
Specifically, they developed a network called ItNet which consists of an iterative loop with
two main components: a pretrained U-net
% for removing artifacts from 128-view FBP
and a data-consistency layer that
enforces agreement between the sinogram of the prediction and the true sinogram.
The pretraining of the U-net is accomplished by training it on a data set that consists
of their own 128-view FBP images paired with the truth images.
% Robust-and-stable perform
%two important tasks to prepare ItNet. First, they estimate the CT system geometry necessary for implementing
%the data-consistency later, and second, the U-net component is pre-trained on artifact mitigation of 128-view FBP images.
The pre-training facilitates the training of the U-nets embedded in the overall ItNet algorithm. The data-consistency
layer has the structure of a least-squares gradient descent step
%\begin{linenomath}
\begin{equation*}
f_{n+1} = f_n -\alpha X^\top(Xf_n-g),
\end{equation*}
%\end{linenomath}
where $n$ is the iteration number and $\alpha$ is a step-size parameter.
In the ItNet implementation, the back-projection is replaced by filtered back-projection
%\begin{linenomath}
\begin{equation*}
f_{n+1} = f_n -\alpha \text{FBP}(Xf_n-g).
\end{equation*}
%\end{linenomath}
Computationally, the method is very efficient;
five iterations of ItNet are used, which entails a total of eleven applications of expensive X-ray projection or back-projection
operations (two operations for each loop plus one back-projection in forming the FBP initial estimate).
There are many other details to the Robust-and-stable strategy, but we highlight only one other point.
Their method also acknowledges the statistical nature of CNN training, and to address this, Robust-and-stable
create an ensemble of 10 trained networks and average their results. 

The second place team, YM\&RH, develop multistage processing for their CNN-based image reconstruction
that also only makes use of the 128-view sinogram data.
As with the previous methodology, YM\&RH estimates the fan-bean scan geometry parameters, and the estimated
forward model is used for both total variation (TV) based image reconstruction and for a variational network.
The first stage involves solving TV-penalized least-squares initialized
with the FBP image. The output of this result fed into a variational network which has an update equation
of the form
%\begin{linenomath}
\begin{equation*}
f_{n+1} = f_n -\alpha X^\top(Xf_n-g) -\beta \text{CNN}(f_n),
\end{equation*}
where $\alpha$ and $\beta$ are the data consistency step length and CNN regularization strength, respectively.
%\end{linenomath}
At the variational network stage, two branches are created. In one branch, seven iterations of the variational
network are followed by two U-net cascades, and in the other branch the variational network is followed
by two attention U-net cascades. The results of the two branches are averaged to yield the final prediction.

The third place team, DEEP\_UL, used staged processing similar to team YM\&RH. They make use of only
the 128-view sinogram data and they estimate the system geometry. DEEP\_UL solves TV-penalized
least-squares in the first stage, but they do so using an ordered-subsets-based algorithm.
For the second stage, they developed their own version of U-net,
which they call high-frequency U-net (HF U-net) to remove the high-frequency artifacts from the output of the first stage.
Similar to Robust-and-stable's implementation, DEEP\_UL used an ensemble of networks and averaged the results to
obtain a final prediction.

The fourth place individual, deepx, is the best performer among all participants who only exploited the 128-view FBP image data.
In this approach,
the exact geometry information is used implicitly as $X^\top$ is used in the process of generating the 128-view FBP image data.
In this approach the CNN processing is used to remove artifacts from the FBP images. The particular network developed
is based on a scale attention network developed for the 2020 MRI brain tumor segmentation challenge. Some modifications
were made to tailor the algorithm to the DL sparse-view CT challenge. This participant also implemented an ensemble
of networks trained with different subsets of the training cases, and used the average to make a final prediction.

The fifth place team, HBB, exploited only the 128-view sinogram data. They
develop a hybrid iterative technique where the trained CNNs are included in the iterative
loop as a regularizer. Their particular design is based on an optimization technique where regularization is performed
on both the image estimate and the corresponding sinogram estimate, and the technique is called joint spatial-Radon (JSR)
domain reconstruction. The JSR optimization problem is solved by the alternating direction method of multipliers (ADMM).
For the CNN-based version, JSR-Net, the sinogram and image regularization update steps include a trained CNN.
The loss function used for training JSR-Net include three terms that measure different aspects of image fidelity:
structural similarity index, similarity in the gradient magnitude image as measured by TV, and mean-square error.
As with approaches one through three, the fan-beam geometry is estimated in order to implement the data consistency step.

These brief overviews only hit some of the highlights of the methodologies and they clearly miss many important
implementation details and parameter settings that are crucial to the excellent performance achieved by these participants.
Readers interested in learning more about any of these methods are encouraged to contact the team members directly.
We point out also that the participating teams are also encouraged to publish their methodology in approaching
the DL-sparse-view CT Challenge.

\section{Discussion and conclusion}
\label{sec:discussion}

The impressive results obtained by the challenge participants motivate a discussion on whether or not DL-based
image reconstruction can solve the sparse-view CT inverse problem. From a practical perspective, there is an
argument that can be made that the error in the top five performing algorithms is negligible. After all the design
of the scan parameters, phantom contrast and image size are representative of what would be encountered in the mid-plane
of an actual breast CT device. Within these conditions, even the worst-case ROI discrepancies do not show visible
error as seen in Fig. \ref{fig:ROIs}; in order to see the ROI discrepancy it is necessary to show the difference images.
Another interesting point is that the first and fourth place teams achieved their high degree of accuracy without
the use of optimization-based methods. Robust-and-stable use only eleven operations that involve forward- or back-projection,
and deepx's image-to-image method is purely based on a CNN. 

Even when discussing practical inversion, there are limitations of the challenge results that mainly have to do with
the set-up of the challenge itself. First, for practical inversion, it needs to be demonstrated that highly accurate
reconstruction can always be obtained reliably. This would require a test set that is much larger than the training set
in order to be convincing, and due to practical considerations of running the challenge we only provided 100 test cases.
Another important issue is that the DL-sparse-view CT challenge uses an object model and scan configuration that we know
is solvable by using optimization-based sparsity-exploiting image reconstruction \cite{sidky2020cnns}. A more convincing
case for DL-based solution to tomography-motivated inverse problems needs to demonstrate solution to a problem where no
other solution exists.

%We now turn back to the original motivating question for the DL-sparse-view CT challenge and discuss what constitutes
%evidence that DL-based image reconstruction can solve the sparse-view CT inverse problem. 
%As solving an inverse
%problem is a pure mathematical question, it needs to be shown that a proposed solution recovers the test phantom
%exactly. Even though the image errors obtained by the winning team are small, they are not zero and strictly
%speaking solution to the sparse-view CT inverse problem cannot be claimed. We do point out, however, that 
%it may be possible to generate numerical evidence that an exact reconstruction principle exists. The study performed
%within the scope of the DL sparse-view CT challenge and the excellent RMSE results obtained by the participants
%is suggestive that such an investigation could be fruitful.
Beyond practical inversion, there is the issue of mathematical solution of inverse problems.
Generating numerical evidence for solving inverse problems in imaging involves studying the image
reconstruction error as a function of various algorithm parameters and investigating what it takes
to achieve a desired RMSE target. Or, put another way, can the image reconstruction be made arbitrarily
accurate. Here are leading issues for such studies:

{\bf Training error:}
Can the algorithm recover the images in the training set exactly? Note that this
study goes against the conventional wisdom of avoiding ``over-training'', where there is a danger
of increasing error on the test set by fitting the training set too closely. For the present set-up
the data are ideal and there is no inconsistency in the training data that would potentially throw off
the NN model fitting. Both the training and testing images need to be recovered accurately.
%The numerical studies for inverse problems question on this particular point focus on how low can the
%training RMSE be made.

{\bf Training set size:}
For the present challenge,
the number of training cases is fixed at 4000. An important extension of the challenge would be to investigate
if the image reconstruction error can be reduced by increasing the training set size.

{\bf Test set size:}
The challenge test set size is fixed at 100 cases. For claiming solution to the CT inverse problem, however,
it is important to demonstrate that the algorithm can reconstruct any image that is representative of the training
data. Specically, for the study in the challenge, it would be important to study the $s_1$ and $s_2$ metrics
as the number of test cases generated from the breast phantom model is increased.

Performing such studies will either reveal limitations of DL-based solution to inverse problems
in imaging or increase confidence that DL-based solution is possible. Aside from the above-mentioned
studies, there are a number of interesting extensions: varying the system size from small toy problems
to full-blown cone-beam CT, testing different phantom models, and investigating different tomographic imaging
modalities.
Even if it turns out that exact recovery by DL-based image reconstruction is not strictly possible, performing
ideal data studies with the intention of reducing RMSE as far as possible can yield important insight into
applying DL-based methods to CT image reconstruction. Each of the above-mentioned study extensions help to
reveal how upper bound performance varies with parameters of the deep-learning algorithm and training set.

These ideal data studies may also advance DL methodology. We point out that even though there is a lot of literature
on deep learning for sparse-view CT image reconstruction, there is a large spread in RMSE results from all
participants. Furthermore, none of the top five performing teams used a published DL-based
algorithm out of the box. Each team designed their own algorithm using ideas from the literature but there were
a number of important adjustments needed in order to achieve the reconstruction accuracy obtained on the test sets.

In terms of general approach for DL-based image reconstruction, the most successful strategy amongst the
participants is to use the sinogram data to find the forward model of the imaging system, then to exploit the
sinogram training data with the forward model. Four of the top five teams use this approach. The ``de-artifacting''
approach using the 128-view FBP images to predict the ground truth images is not as successful in reducing RMSE.
One team, however, did rank in the top five using this approach. For either approach the large spread in results
hints at the possibility that further reduction in test RMSE can be achieved for the sparse-view CT system specified
in the challenge. Further studies along this line can help to clarify the use of deep-learning for solving
inverse problems in imaging.

\section*{Acknowledgment}
We are deeply in debt to the AAPM Working Group on Grand Challenges (WGGC):
Sam Armato,
Kenny Cha,
Karen Drukker,
Keyvan Farahani,
Lubomir  Hadjiyski,
Reshma Munbodh,
Nicholas Petrick, and
Emily Townley.
The WGGC accepted our challenge proposal, connected us with valuable resources namely
the MedICI challenge platform, and guided us through the details of setting up a Medical Imaging Grand Challenge.
We had outstanding support from
Benjamin Bearce and Jayashree Kalpathy-Cramer at the MedICI challenge platform for the actual
implementation of the challenge website and for addressing any technical issues that came up during the challenge.
Of course, we are also extremely grateful to the
participants who took the time to engage in the DL-sparse-view CT challenge in a meaningful way and from whom
we have learned a lot about applying DL to sparse-view CT image reconstruction.
In particular, we thank the test phase participants who submitted reports describing their algorithms (in alphabetical order):
Laslo van Anrooij,
Navchetan Awasthi,
R\'emy B\'edard,
C\'edric B\'elanger,
Seungryong Cho,
Tijian Deng,
Alexander Denker,
Bin Dong,
Congcong Du,
Xiaohong Fan,
Martin Genzel,
Daniel Gourdeau,
Su Han,
Yu Hu,
Mei Huang,
Andreas Hauptmann,
Reinhard Heckel,
Satu I. Inkinen,
Hyeongseok Kim,
Andreas Kofler,
Maxence Larose,
Seoyoung Lee,
Johannes Leuschner,
Gang Li,
Ling Li,
Yu Li,
Baodong Liu,
Qian Liu,
Ke Lu,
Yang Lu,
Jan Macdonald,
Youssef Mansour, 
Maximillian M\"arz,
Peizhang Qian,
Zhiwei Qiao,
Leonardo Di Schiavi Trotta,
Maximillian Schmidt,
Hailai Tian,
Yin Yang,
Yading Yuan,
Haimiao Zhang,
Jianping Zhang,
Yanjiao Zhang, and
Bo Zhou.
This work is supported in part by NIH
Grant Nos. R01-EB026282 and R01-EB023968.
The contents of this article are solely the responsibility of
the authors and do not necessarily represent the official
views of the National Institutes of Health.

%\bibliographystyle{medphy}
%\bibliography{cnntv}
\section*{Conflict of interest}
The authors have no conflicts to disclose.

\section*{Data availability}
The DL-sparse-view CT challenge data are available at the challenge website:
\url{https://dl-sparse-view-ct-challenge.eastus.cloudapp.azure.com/competitions/1}.
To access the data, please sign up for the challenge and then follow the instructions
under the ``Participate'' tab.

\end{document}